\documentclass [10.5pt]{article}
\usepackage{amsfonts,mathrsfs}
\usepackage{amsmath,amssymb}
\usepackage{cite}
\usepackage{graphicx}
\usepackage{psfrag}

\begin{document}
\title{Constructions of Almost Optimal Resilient Boolean Functions on Large Even Number of Variables
\thanks{Published in IEEE Transactions on Information Theory, vol. 55, no. 12, 2009.
 (doi: 10.1109/TIT.2009.2032736)}}
 \author{WeiGuo ZHANG\thanks{e-mail: w.g.zhang@qq.com}~ and GuoZhen XIAO
\\ISN Lab, Xidian University, Xi'an 710071, P.R.China
}
\date{}

\maketitle

\begin{abstract}
In this paper, a technique on constructing nonlinear resilient
Boolean functions is described. By using several sets of disjoint
spectra functions on a small number of variables, an almost optimal
resilient function on a large even number of variables can be
constructed. It is shown that given any $m$, one can construct
infinitely many $n$-variable ($n$ even), $m$-resilient functions
with nonlinearity $>2^{n-1}-2^{n/2}$. A large class of highly
nonlinear resilient functions which were not known  are obtained.
Then one method to optimize the degree of the constructed functions
is proposed. Last, an improved version of the main construction is
given.

\end{abstract}


\section{Introduction}
Boolean functions are used as nonlinear combiners or nonlinear
filters in certain models of stream cipher systems. In the design of
cryptographic Boolean functions, there is a need to consider a
multiple number of criteria simultaneously. The widely accepted
criteria are balancedness, high nonlinearity, high algebraic degree,
and correlation immunity of high order (for balanced functions,
correlation immunity is referred to as resiliency).

By an $(n, m, d, N_f)$ function we mean an $n$-variable,
$m$-resilient (order of resiliency $m$) Boolean function $f$ with
algebraic degree $d$ and nonlinearity $N_f$.

Unfortunately, all the criteria above cannot be maximized together.
For $n$ even, the most notable example is perhaps bent functions
\cite{Rothaus}. Achieving optimal nonlinearity
$2^{n-1}-2^{{n/2}-1}$, bent functions permit to resist linear
attacks in the best possible way. But they are improper for
cryptographic use because they are neither balanced nor
correlation-immune and their algebraic degrees are not more than
$n/2$. When concerning the order of resiliency, Siegenthaler
\cite{Siegenthaler} and Xiao \cite{Xiao} proved that $d\leq n-m-1$
for $m$-reslient Boolean functions. Such a function, reaching this
bound, is called degree-optimized.

For the reasons above, it is more important to construct those
degree-optimized resilient Boolean functions which have almost
 optimal (large but not optimal) nonlinearity, say between $2^{n-1}-2^{n/2}$
and $2^{n-1}-2^{{n/2}-1}$, when $n$ is even. This is also what we do
in this paper.

We now give a summary of earlier results that are related to our
work.

1)  To obtain nonlinear resilient functions, a modification of the
Maiorana-McFarland (M-M) construction of bent functions (cf.
\cite{Dillon}) by concatenating the small affine functions was first
employed by Camion et al \cite{Camion} and later studied in \cite
{Seberry-e}, \cite{Chee}, \cite{Sarkar-e}. The nonlinearity of
$n$-variable M-M
 resilient functions cannot exceed $2^{n-1}-2^{\lfloor
n/2 \rfloor}$. The M-M technique in general does not generate
degree-optimized functions and the M-M functions are potentially
cryptographically weak \cite{Carlet},\cite{Khoo}.

2) An interesting extension of the M-M technique has been made by
Carlet \cite {Carlet}, where the concatenation of affine functions
is replaced by concatenation of quadratic functions.  In general,
these constructed functions can not be degree-optimized, and the
other parameters such as nonlinearity and resiliency are not better
than those of the M-M functions.

3) Pasalic \cite{Pasalic} presented a revised version of the M-M
technique to obtain degree-optimized resilient functions. The
modification is simple and smart but the nonlinearity value of the
constructed functions is at most $2^{n-1}-2^{\lfloor n/2 \rfloor}$.

4) Sarkar and Maitra \cite{Sarkar-e} indicated that for each order
of resiliency $m$, it is possible to find an even positive integer
$n$ to construct an $(n, m, n-m-1, N_f)$ function $f$ with
$N_f>2^{n-1}-2^{ n/2 }$. They showed that for even $n\geq 12$, the
nonlinearity of 1-resilient functions with maximum possible
algebraic degree  $n-2$ can reach
$2^{n-1}-2^{n/2-1}-2^{n/2-2}-2^{n/4-2}-4$. It was further improved
due to the method proposed by Maitra and Pasalic \cite{Maitra-IEEE}.
Thanks to the existence of the $(8, 1, 6, 116)$ functions, an $(n,
1, n-2, N_f)$ function $f$ with $N_f=2^{n-1}-2^{n/2-1}-2^{n/2-2}-4$
could be obtained, where $n\geq 10$.

5) Seberry et al. \cite{Seberry-c} and Dobbertin \cite{Dobbertin}
independently presented constructions of highly nonlinear balanced
Boolean functions by modifying the M-M bent functions. To obtain an
$n$-variable balanced function, they concatenated $2^{n/2}-1$
nonconstant distinct $n/2$-variable linear functions and one
$n/2$-variable modified M-M class highly nonlinear balanced Boolean
function which can be constructed in a recursive manner. These
constructed functions attain the best known nonlinearity for
$n$-variable ($n$ even) balanced functions. Unfortunately, these
functions are not 1-resilient functions.

6) To obtain $m$-resilient functions with nonlinearity
$>2^{n-1}-2^{n/2-1}-2^{n/2-2}$ for $n$ even and $n\geq14$, Maitra
and Pasalic \cite{Maitra-DAM} applied the concatenation of
$2^{n/2}-2^k$ distinct linear $m$-resilient functions on $n/2$
variables together with a highly nonlinear resilient function on
$n/2+k$ variables. Moreover, they have provided a generalized
construction method for $m$-resilient functions with nonlinearity
 $2^{n-1}-2^{n/2-1}-2^{n/2-3}-2^{n/2-4}$ for all $n\geq 8m+6$.
For sufficiently large $n$, it is possible to get such functions
with nonlinearity $\approx 2^{n-1}-2^{n/2-1}-\frac{2}{3}2^{n/2-2}$.
And it is the upper bound on maximum possible nonlinearity under
their construction method.

7) Computer search techniques have played an important role in the
design of cryptographic Boolean functions in the last ten years
\cite{Millan-i}, \cite{Millan-e}, \cite{Millan-a}, \cite{Clark}. For
a small number of variables, Boolean functions with good
cryptographic parameters could be found by using heuristic search
techniques \cite{Maitra-IEEE}, \cite{Saber}, \cite{Kavut}. However,
search techniques cannot be used for functions with a large number
of variables at present.

8) During the past decade, the most infusive results on the design
of cryptographic Boolean functions were centered on finding small
functions with desirable cryptographic properties. When it comes to
constructing large functions, people used recursive constructions
\cite{Tarannikov}, \cite{Tarannikov-FSE}, \cite{Fedorova} besides
the M-M construction and its revised (extended) versions. With the
rapid development of integrated circuit technique, Boolean functions
with large number of variables can be easily implemented in hardware
\cite{Sarkar-ieee}.

In this paper we propose a technique to construct high nonlinear
resilient Boolean functions on large even number of variables
$(n\geq 12)$. We obtain a large class of resilient Boolean functions
with a nonlinearity higher than that attainable by any previously
known construction method.

The organization of this paper is as follows. In Section II, the
basic concepts and notions are presented. In Section III, we present
a method to construct a set of ``disjoint spectra functions" by
using a class of ``partially linear functions". Our main
construction is given in Section IV. A method for constructing
resilient functions on large even number of input variables is
proposed. We show that all the constructed functions are almost
optimal. In Section V, the degrees of the constructed functions are
optimized. In Section VI, an improved version of the main
construction is given. Finally, Section VII concludes the paper with
an open problem.

\section{Preliminary}

To avoid confusion with the additions of integers in $\mathbb{R}$,
denoted by $+$ and $\Sigma_i$, we denote the additions over
$\mathbb{F}_2$ by $\oplus$ and $\bigoplus_i$. For simplicity, we
denote by $+$ the addition of vectors of $\mathbb{F}_2^n$. A Boolean
function of $n$ variables is a function from $\mathbb{F}_2^n$ into
$\mathbb{F}_2$, and we denote by $\mathcal {B}_n$ the set of all
Boolean functions of $n$ variables. A Boolean function
$f(X_n)\in\mathcal {B}_n$, where $X_n=(x_1, \cdots, x_n)\in
\mathbb{F}_2^n$, is generally represented by its algebraic normal
form (ANF)
\begin{equation}
 f(X_n)=\bigoplus_{u\in\mathbb{F}_2^n}\lambda_u(\prod_{i=1}^n
 x_i^{u_i})
\end{equation}
 where $\lambda_u\in\mathbb{F}_2$ and $
 u=(u_1,\cdots,u_n).$
The algebraic degree of $f(X_n)$, denoted by $deg(f)$, is the
maximal value of $wt(u)$ such that $\lambda_u\neq 0$, where $wt(u)$
denotes the Hamming weight of $u$. A Boolean function with
$deg(f)\leq 1$ is said to be affine. In particular, an affine
function with constant term equal to zero is called a linear
function.  Any linear function on $\mathbb{F}_2^n$ is denoted by
\begin{equation*}
\omega \cdot X_n=\omega_1x_1\oplus\cdots \oplus \omega_nx_n
\end{equation*}
where $\omega=(\omega_1,\cdots,\omega_n)\in \mathbb{F}_2^n$. The
Walsh spectrum of $f\in \mathcal {B}_n$ in point $\omega$ is denoted
by $W_f(\omega)$ and calculated by
\begin{equation}
 W_f(\omega)=\sum_{X_n\in \mathbb {F}_2^n} (-1)^{f(X_n)\oplus \omega\cdot X_n}.
\end{equation}
$f\in \mathcal {B}_n$ is said to be balanced if its output column in
the truth table contains equal number of $0$'s and $1$'s (i.e.
$W_f(0)=0$).

In \cite{Xiao}, a spectral characterization of resilient functions
has been presented.

\emph{Lemma 1:} An $n$-variable Boolean function is $m$-resilient if
and only if its Walsh transform satisfies
\begin{equation}\label{3}
W_f(\omega)=0, \textrm{ for $0\leq wt(\omega)\leq m$, $\omega\in
\mathbb{F}_2^n$}.
\end{equation}

The Hamming distance between two $n$-variable Boolean functions $f$
and $\rho$ is denoted by
$$d(f,\rho)=\{X_n\in \mathbb{F}_2^n: f(X_n)\neq \rho(X_n)\}.$$
The set of all affine functions on $\mathbb{F}_2^n$ is denoted by
$A(n)$. The nonlinearity of a Boolean function $f\in \mathcal {B}_n$
is its distance to the set of all affine functions and is defined as
$$N_f=\min_{\rho\in A(n)}(d(f,\rho)).$$ In term of Walsh spectra, the
nonlinearity of $f$ is given by \cite{Merer}
\begin{equation}\label{4}
N_f=2^{n-1}-\frac{1}{2}\cdot \max_{\omega\in
\mathbb{F}_2^n}|W_f(\omega)|.
\end{equation}
 Parseval's equation \cite{MacWilliams}
states that
\begin{equation}
\sum_{\omega\in \mathbb{F}_2^n}(W_f(\omega))^2=2^{2n}
\end{equation}
and implies that
\begin{equation*}
N_f\leq 2^{n-1}-2^{n/2-1}.
\end{equation*}
The equality occurs if and only if $f\in \mathcal {B}_n$ are bent
functions, where $n$ is even.

Bent functions can be constructed by the M-M method. The original
M-M functions are defined as follows: for any positive integers $p$,
$q$ such that $n=p+q$, an M-M function is a function $f\in \mathcal
{B}_n$ defined by
\begin{equation}\label{}
f(Y_q, X_p)=\phi(Y_q)\cdot X_p\oplus \pi(Y_q), ~~ X_p\in
\mathbb{F}_2^p, Y_q\in \mathbb{F}_2^q
\end{equation}
where $\phi$ is any mapping from $\mathbb{F}_2^q$ to
$\mathbb{F}_2^p$ and $\pi\in \mathcal {B}_q$. When $n$ is even,
$p$=$q$=$n/2$, and $\phi$ is injective, the M-M  functions are bent.
Certain choices of $\phi$ can easily yield bent functions with
degree $n/2$.  For the case of $n=2$, $f\in \mathcal {B}_2$ is bent
if and only if $deg(f)=2$.

The M-M construction is in essence a concatenation of affine
functions. The following definition shows a more general approach to
obtain a ``large"  Boolean function by concatenating the truth
tables of any small Boolean functions.

\emph{Definition 1:} Let $Y_q\in \mathbb{F}_2^q$, $X_p\in
\mathbb{F}_2^p$, and $p$, $q$ be positive numbers with $p+q=n$.
$f\in \mathcal {B}_n$ is called a concatenation of the functions in
the set $G=\{g_b\ |\ b\in \mathbb{F}_2^q \}\subset \mathcal {B}_p$
if
\begin{eqnarray}
f(Y_{q},X_{p})= \bigoplus_{b\in \mathbb{F}_2^{q}}Y_{q}^b\cdot
g_b(X_{p}),
\end{eqnarray}
where the notation $Y_{q}^b$ is defined  by
\begin{eqnarray}\label{8}
Y_{q}^b = \left\{ \begin{array}{ll}
1 & \textrm{if $Y_{q}=b$}\\
0 & \textrm{if $Y_{q}\neq b$}.
\end{array} \right.
\end{eqnarray}

Theorem 2 in \cite{Siegenthaler} allows us to verify that the
following lemma is true.

\emph{Lemma 2:} With the same notation as in Definition 1, if all
the functions in $G$ are $m$-resilient functions, then $f$ is an
$m$-resilient function.

From now on, we will focus on highly nonlinear resilient Boolean
functions with an even number of variables in the following sense.

\emph{Definition 2:} Let $n\geq 4$ be even. $f\in \mathcal {B}_n$ is
said to be almost optimal if
\begin{eqnarray}
2^{n-1}-2^{n/2}\leq N_f < 2^{n-1}-2^{n/2-1}.
\end{eqnarray}

\section{A Large Set of Disjoint Spectra Functions}

Disjoint spectra functions will play an important role in
constructing almost optimal resilient functions in this paper.

 \emph{Definition 3:} A set of Boolean functions $\{g_1, g_2,
\cdots, g_e \}\subset \mathcal {B}_p$ such that for any $\alpha\in
\mathbb{F}_2^p$,
\begin{eqnarray}\label{dsf}
W_{g_i}(\alpha)\cdot W_{g_j}(\alpha)=0, \ \ 1\leq i < j \leq e
\end{eqnarray}
is called a set of disjoint spectra functions.

The idea that two Boolean functions with disjoint spectra can be
used to construct highly nonlinear resilient functions was clearly
mentioned in \cite{Pasalic-wcc}, and it was also used in
\cite{Tarannikov}, \cite{Fedorova}, \cite{Maitra-IEEE}. In this
section, we provide a simple construction method for a large set of
disjoint spectra functions by using a set of ``partially linear"
functions.

As a family of special resilient functions, partially linear
functions were firstly considered by Siegenthaler
\cite{Siegenthaler}. Here is the definition of such functions.

{\emph{Definition 4:}} Let $t$ be a positive integer and $\{i_1,
\cdots, i_t\}\cup \{i_{t+1}, \cdots, i_p\}=\{1, \cdots, p\}$. Let
$X_p=(x_1, \cdots, x_p)\in \mathbb{F}_2^p$, $X'_t=(x_{i_1}, \cdots,
x_{i_{t}})\in \mathbb{F}_2^t$ and $X''_{p-t}=(x_{i_{t+1}}, \cdots,
x_{i_{p}})\in \mathbb{F}_2^{p-t}$. For any $c\in \mathbb{F}_2^t$,
$g_c\in \mathcal {B}_p$ is called a $t$th-order partially linear
function if
\begin{eqnarray}
g_c(X_p)=c\cdot X'_t \oplus h_c(X''_{p-t})
\end{eqnarray} where $h_c\in
\mathcal {B}_{p-t}$.

Now we use partially linear functions to construct a set of disjoint
spectra functions.

{\emph{Lemma 3:}}  With the same notation as in Definition 4, a set
of $t$th-order partially linear functions
\begin{eqnarray}
T=\{g_c(X_p)=c\cdot X'_t\oplus h_c(X''_{p-t}) ~|~c\in
\mathbb{F}_2^t\}
\end{eqnarray}
is a set of disjoint spectra functions.

{ \emph{Proof:}} Let $\alpha=(\delta,\theta)\in \mathbb{F}_2^{p}$,
where $\delta \in \mathbb{F}_2^{t}$ and $\theta \in
\mathbb{F}_2^{p-t}$. For any $g_c\in T$,
\begin{align}\label{}
W_{g_c}(\alpha)&= \sum_{X_{p}\in \mathbb{F}_2^{p}}(-1)^{c \cdot X'_t\oplus h_c(X''_{p-t})\oplus \alpha \cdot X_{p}}\nonumber\\
&= \sum_{X_{p} \in \mathbb{F}_2^{p}}(-1)^{(c+\delta) \cdot X'_t\oplus (h_c(X''_{p-t})\oplus \theta \cdot X''_{p-t})}\nonumber\\
&= \sum_{X'_t\in \mathbb{F}_2^{t}}(-1)^{(c+\delta) \cdot
X'_t}\sum_{X''_{p-t} \in
\mathbb{F}_2^{p-t}}(-1)^{(h_c(X''_{p-t})\oplus \theta \cdot X''_{p-t})}\nonumber\\
&= \left(\sum_{X'_t\in \mathbb{F}_2^{t}}(-1)^{(c+\delta) \cdot
X'_t}\right)\cdot W_{h_c}(\theta)
\end{align}
We have
\begin{eqnarray}\label{wgc}
W_{g_c}(\alpha) = \left\{ \begin{array}{ll}
0 & \textrm{if $c\neq \delta$}\\
2^t\cdot W_{h_c}(\theta) & \textrm{if $c = \delta$}.
\end{array} \right.
\end{eqnarray}
For any $g_{c'}\in T$, $c'\neq c$, we have
$$W_{g_c}(\alpha)\cdot W_{g_{c'}}(\alpha)=0.$$
According to Definition 3, $T$ is a set of disjoint spectra
functions.

Disjoint spectra functions (partially linear functions) will be used
as the ``components" to construct almost optimal resilient Boolean
functions in this paper.

{\emph{Open Problem:}} Construct a large set of  disjoint spectra
functions which are not (linearly equivalent to) partially linear
functions.

\section{Main Construction}

This section presents a method for constructing resilient Boolean
functions with very high nonlinearity. The algebraic degrees of the
functions are also given.

 {\emph{Construction 1:}} Let $n\geq 12$ be an  even number, $m$ be a positive number, and $(a_1, \cdots, a_s)\in \mathbb{F}_2^s$ such that
\begin{equation}\label{constrction 1}
\sum_{j=m+1}^{n/2} {{n/2} \choose j}+\sum_{k=1}^s\left(a_k\cdot
\sum_{j=m+1}^{n/2-2k} {{n/2-2k} \choose j}\right)\geq 2^{n/2}
\end{equation} where
$s=\lfloor(n-2m-2)/4\rfloor$.  Let  $X_{n/2}=(x_1,\cdots,
x_{n/2})\in \mathbb{F}_2^{n/2}$, $X'_{t}=(x_1,\cdots, x_{t})\in
\mathbb{F}_2^{t}$, and $X''_{2k}=(x_{t+1},\cdots, x_{n/2})$ $\in
\mathbb{F}_2^{2k}$ with $t+2k=n/2$. Let
\begin{equation}
\Gamma_0=\{c \cdot X_{n/2}\ | \ c\in \mathbb{F}_2^{n/2}, \ wt(c)>m
\}.
\end{equation}
For $1\leq k \leq s$, let $H_k$ be a nonempty set of $2k$-variable
bent functions with algebraic degree $\max (k, 2)$ and
\begin{align}
\Gamma_k=\{c \cdot X'_t\oplus h_c(X''_{2k}) ~| ~ c\in
\mathbb{F}_2^t, wt(c)>m\}
\end{align}
where $h_c\in H_k$.  Set
\begin{eqnarray}
\Gamma=\bigcup_{k=0}^s\Gamma_k.
\end{eqnarray}
Denote by $\phi$ any injective mapping from $\mathbb{F}_2^{n/2}$ to
$\Gamma$. Then for $(Y_{n/2}, X_{n/2})\in \mathbb{F}_2^{n/2}\times
\mathbb{F}_2^{n/2}$ we construct the function $f\in \mathcal {B}_n$
as follows:
\begin{eqnarray}
f(Y_{n/2},X_{n/2})= \bigoplus_{b\in
\mathbb{F}_2^{n/2}}Y_{n/2}^b\cdot \phi(b).
\end{eqnarray}

\emph{Remark 1:}

1) For Inequality (\ref{constrction 1}) holds, we have
\begin{eqnarray}
|\Gamma|=\sum_{k=0}^s |\Gamma_k|\geq 2^{n/2}.
\end{eqnarray}
Due to this we can find an injective mapping $\phi$.

2) All the functions in $\Gamma$ are partially linear functions and
each $\Gamma_k$ is a set of disjoint spectra functions.

 { \emph{Theorem 1:}} Let $f\in \mathbb{F}_2^n$ be as in Construction 1. Then $f$ is an almost optimal
$(n,m,d,N_f)$ function with
\begin{equation} \label{21}
N_f\geq 2^{n-1}-2^{n/2-1}-\sum_{k=1}^s (a_k\cdot 2^{n/2-k-1})
\end{equation}
 and
\begin{equation}
d\leq n/2+\max \{2, \max \{k ~ | ~ a_k\neq 0, ~ k=1,2, \cdots s\}\}.
\end{equation}

 { \emph{Proof:}}
 For any $(\beta,\alpha)\in \mathbb{F}_2^{n/2}\times
\mathbb{F}_2^{n/2}$ we have
\begin{align}
W_f(\beta,\alpha)&=\sum_{(Y_{n/2},X_{n/2})\in
\mathbb{F}_2^n}(-1)^{f(Y_{n/2}, X_{n/2})\oplus (\beta,\alpha)\cdot (Y_{n/2}, X_{n/2})} \nonumber\\
&= \sum_{b\in \mathbb{F}_2^{n/2}}(-1)^{\beta\cdot b}\sum_{X_{n/2}\in
\mathbb{F}_2^{n/2} }(-1)^{g_b(X_{n/2})\oplus \alpha\cdot X_{n/2}} \nonumber\\
&= \sum_{b\in \mathbb{F}_2^{n/2}}(-1)^{\beta\cdot
b}W_{g_b}(\alpha) \nonumber\\
&= \sum_{k=0}^s \sum_{\phi(b)\in \Gamma_k \atop b\in
\mathbb{F}_2^{n/2}}(-1)^{\beta\cdot b}W_{g_b}(\alpha) \label{23}
\end{align}
Let $0\leq k\leq s$. Any $g_b\in \Gamma_k$ is a partially linear
function. From (\ref{wgc}), we have
$$
W_{g_b}(\alpha)\in \{0, \pm 2^{n/2-k}\}.
$$
Let
\begin{eqnarray}
A_k=\Gamma_k \cap \{\phi(b) \ | \  b\in \mathbb{F}_2^{n/2} \}.
\end{eqnarray}
Since (\ref{constrction 1}) holds, there exists an injective mapping
$\phi$ such that
\begin{eqnarray} \label{25}
\sum_{k=1}^s  |A_k|=\sum_{i=1}^m {{n/2} \choose i}.
\end{eqnarray}
From Lemma 3, $\Gamma_k$ is a set of disjoint spectra functions.
Noting (\ref{dsf}), if $A_k\neq \emptyset,$ then we have
\begin{eqnarray}\label{26}
\sum_{\phi(b)\in \Gamma_k \atop b\in
\mathbb{F}_2^{n/2}}(-1)^{\beta\cdot b}W_{g_b}(\alpha)\in \{0, \pm
2^{n/2-k}\}.
\end{eqnarray}
If $A_k=\emptyset,$ then we have
\begin{eqnarray}
\sum_{\phi(b)\in \Gamma_k \atop b\in
\mathbb{F}_2^{n/2}}(-1)^{\beta\cdot b}W_{g_b}(\alpha)=0. \label{27}
\end{eqnarray}
Combining (\ref{23}), (\ref{26}), and (\ref{27}), we have
\begin{eqnarray}\label{28}
|W_f(\beta,\alpha)|\leq 2^{n/2}+\sum_{k=1}^s a_k\cdot 2^{n/2-k}
\end{eqnarray}
 where
\begin{eqnarray}
a_k = \left\{ \begin{array}{ll}
0 & \textrm{if $A_k=\emptyset$}\\
1 & \textrm{if $A_k\neq \emptyset$}.
\end{array} \right.
\end{eqnarray}
From (\ref{4}), Inequality (\ref{21}) holds. From Definition 2, $f$
is almost optimal.

Note that the algebraic degree of any function in $\Gamma_1$ is 2.
Hence, when
\begin{eqnarray}
\max \{k ~ | ~a_k\neq 0, \ k=1,2, \cdots s\}=1
\end{eqnarray}
$d\leq n/2+2$  where the equality holds if and only if $|A_1|$ is
odd. Note that the algebraic degree of any bent functions on
$\mathbb{F}_2^{2k}$ can reach $k$ when $k\geq 2$. So $d$ can reach
$n/2+k'$ when $k'\geq 2$ and $|A_{k'}|$ is odd, where
\begin{eqnarray}
k'=\max \{k \ | \ a_k\neq 0, \ k=1,2, \cdots s\}.
\end{eqnarray}

Any $g_b\in \Gamma$, $b\in \mathbb{F}_2^{n/2}$, is an $m$-resilient
function, where $g_b(X_{n/2})=\phi(b)$. From Lemma 2,  $f$ is an
$m$-resilient function since it is a concatenation of $m$-resilient
functions.  $\Box$

{\emph{Remark 2:}}

1) The nonlinearity of the resilient functions constructed above is
always strictly greater than $2^{n-1}-2^{n/2}$. For reasonable fixed
$n$ and $m$, the nonlinearity of the constructed functions is always
greater than that of the known ones except some functions on small
even number of variables.

2) Let $m$ be the maximum number such that Inequality
(\ref{constrction 1}) holds. Roughly speaking, $m/n$ tends to 1/4.

{ \emph{Example 1:}} It is possible to construct a $(16, 1, 10,
2^{15}-2^7-2^5)$ function.

Note that $s=\lfloor (n-2m-2)/4\rfloor$=3. For $1\leq k\leq 3$, let
$X'_{8-2k}=(x_1, \cdots, x_{8-2k})\in \mathbb{F}_2^{8-2k}$ and
$X''_{2k}=(x_{8-2k+1}, \cdots, x_{8})\in \mathbb{F}_2^{2k}$. Let
$X_{8}=(X'_{8-2k}, X''_{2k})\in \mathbb{F}_2^{8}$. We construct four
sets of disjoint spectra functions as follows:
$$\Gamma_0=\{c\cdot X_{8} \ | \ wt(c)>1, c\in
\mathbb{F}_2^{8}\}.$$ For $1\leq k\leq 3$,
$$\Gamma_k=\{c\cdot X'_{8-2k} \oplus h_c(X''_{2k}) \ | \ wt(c)>1, c\in
\mathbb{F}_2^{8-2k}\}$$ where $h_c\in H_{k}$. We have
$$|\Gamma_k|=\sum_{i=2}^{8-2k}{{8-2k} \choose i}, \ 0\leq k\leq 3.$$
Notice that
$$|\Gamma_0|+|\Gamma_2|=258 > 2^{8},$$
it is possible to establish an injective mapping $\phi$ from
$\mathbb{F}_2^{8}$ to $\Re$, where $\Re=\Gamma_0 \cup \Gamma_2$.
Then for $(Y_{8}, X_{8})\in \mathbb{F}_2^{8}\times \mathbb{F}_2^{8}$
we construct the function $f\in \mathcal {B}_{16}$ as follows:
$$f(Y_{8}, X_{8})= \bigoplus_{b\in \mathbb{F}_2^{8}}Y_{8}^b\cdot \phi(b).$$
From (\ref{28}), for any $(\beta, \alpha)\in \mathbb{F}_2^{8}\times
\mathbb{F}_2^{8}$, we have
\begin{align*}
 \max_{(\beta, \alpha)\in \mathbb{F}_2^{8}} |W_f(\beta, \alpha)|
\leq \sum_{k=0}^3 \sum_{g\in \Gamma_k}|W_g(\alpha)| =
\sum_{k=0}^3\max_{\alpha\in \mathbb{F}_2^{8}\atop {g\in
\Gamma_k}}|W_g(\alpha)| =2^{8}+2^{6}
\end{align*}
By (\ref{4}), we have
$$N_f\geq 2^{15}-2^{7}-2^{5}.$$ Note that the partially linear
function in $\Gamma_2$ can be denoted by $g=c\cdot X'_{4} \oplus
h_c(X''_{4})$ where $h_c$ is a bent function on $\mathbb{F}_2^4$.
Since the algebraic degree of $h_c(X''_{4})$ can reach 2, $deg(f)$
can reach 8+2=10. So it is possible to obtain a $(16, 1, 10,
2^{15}-2^{7}-2^{5})$ function.

\section{Degree Optimization}

Let
\begin{eqnarray*}
\{i_1, \cdots, i_{m+1}\}\cup \{i_{m+2},  \cdots, i_{n/2}\}=\{1,
\cdots,n/2\}.
\end{eqnarray*}
The algebraic degree of any $(n,m,d,N_f)$ function $f$ obtained in
Construction 1 can be optimized by adding a monomial
$x_{i_{m+2}}\cdots x_{i_{n/2}}$ to one function $g\in \Gamma$ with
$\phi^{-1}(g)\neq \emptyset$, where $g$ can be denoted by
\begin{align}
g=x_{i_1}\oplus\cdots \oplus x_{i_{m+1}}\oplus
\hbar(x_{i_{m+2}},\cdots, x_{i_{n/2}}).
\end{align}
It is not difficult to prove that $N_{f'}\in \{N_f, N_f-2^{m+1}\}$,
where $N_{f'}$ is the nonlinearity of the degree-optimized function
$f'$. To optimize the algebraic degree of $f$ and ensure that
$N_{f'}=N_f$, we below propose an idea to construct a set of
disjoint spectra functions $\Gamma'_0$ including a nonlinear
function $g'=g + x_{i_{m+2}}\cdots x_{i_{n/2}}$.

{\emph{Construction 2:}} Let $n\geq 12$ be an  even number, $m$ be a
positive number, and $(a_1, \cdots, a_s)\in \mathbb{F}_2^s$ such
that
\begin{align}
\left(\sum_{j=m+1}^{n/2} {{n/2}\choose j}-2^{n/2-m-1}+1\right)
+\sum_{k=1}^s\left(a_k \cdot \sum_{j=m+1}^{n/2-2k} {{n/2-2k}\choose
j}\right)\geq 2^{n/2}
\end{align}
where $s=\lfloor(n-2m-2)/4\rfloor$. Let
\begin{align}
S= \{c~|~c=(c_{1},\cdots, c_{n/2})\in \mathbb{F}_2^{n/2}, wt(c)>m,
(c_{i_1}, \cdots, c_{i_{m+1}})=(1\cdots1)\}.
\end{align}
 Let
\begin{eqnarray}
g'(X_{n/2})=c' \cdot X_{n/2}\oplus x_{i_{m+2}}x_{i_{m+3}}\cdots
x_{i_{n/2}}
\end{eqnarray} where $c'\in
S$. For $1\leq k\leq s$, $\Gamma_k$ is defined as in Construction 1.
And we modify the $\Gamma_0$ in Construction 1 as follows:
\begin{align}\label{36}
\Gamma_0'=\{g'(X_{n/2})\} \cup  \{c \cdot X_{n/2}
   |   c\in \mathbb{F}_2^{n/2},  wt(c)>m,c\notin S \}.
\end{align}
Set
\begin{eqnarray}
\Gamma'=\Gamma'_0\cup\Gamma_1\cup\cdots \cup\Gamma_s.
\end{eqnarray}
Denote by $\phi'$ any injective mapping from $\mathbb{F}_2^{n/2}$ to
$\Gamma'$ such that $\phi'^{-1}(g_{c'})\neq \emptyset$. The function
$f'\in \mathcal {B}_n$ is constructed as follows:
\begin{eqnarray}
f'(Y_{n/2},X_{n/2})= \bigoplus_{b\in
\mathbb{F}_2^{n/2}}Y_{n/2}^b\cdot \phi'(b).
\end{eqnarray}

{\emph{Theorem 2:}} The function $f'\in \mathbb{F}_2^n$ proposed by
Construction 2 is an almost optimal $(n,m,n-m-1,N_{f'})$ function
with
\begin{eqnarray}
N_{f'}\geq 2^{n-1}-2^{n/2-1}-\sum_{k=1}^s a_k\cdot 2^{n/2-k-1}.
\end{eqnarray}

{ \emph{Proof:}} $f'$ is an $m$-resilient function since it is a
concatenation of $m$-resilient functions.

Let $g_b(X_{n/2})=\phi'(b)\in \Gamma'$, $b\in \mathbb{F}_2^{n/2}$.
From the proof of Theorem 1, for any $(\beta, \alpha)\in
\mathbb{F}_2^{n/2}\times \mathbb{F}_2^{n/2}$, we have
\begin{align}
W_{f'}(\beta, \alpha) =\sum_{\phi(b)\in \Gamma'_0 \atop b\in
\mathbb{F}_2^{n/2}}(-1)^{\beta\cdot b}W_{g_b}(\alpha) +\sum_{k=1}^s
\sum_{\phi(b)\in \Gamma_k \atop b\in
\mathbb{F}_2^{n/2}}(-1)^{\beta\cdot b}W_{g_b}(\alpha).
\end{align}
Let  $c'=(c'_{1}, \cdots, c'_{n/2})\in \mathbb{F}_2^{n/2}$ and
$\alpha=(\alpha_1,\cdots, \alpha_{n/2})\in \mathbb{F}_2^{n/2}.$ We
have
\begin{align}
W_{g'}(\alpha)
&= \sum_{X_{n/2} \in \mathbb{F}_2^{n/2}}(-1)^{ (c'+\alpha) \cdot X_{n/2}  \oplus x_{i_{m+2}}\cdots~ x_{i_{n/2}}}\nonumber \\
&= \left\{ \begin{array}{ll}
2^{n/2}-2^{m+2}& \textrm{if $\alpha=c'$}\\
\pm 2^{m+2} & \textrm{if $\alpha\neq c'$ and $\theta= \delta$}\\
0 & \textrm{if $\alpha\neq c'$ and $\theta\neq\delta$}.
\end{array} \right.
\end{align}
where $\delta=(c'_{i_{1}}, \cdots, c'_{i_{m+1}})$ and
$\theta=(\alpha_{i_{1}}, \cdots, \alpha_{i_{m+1}})$. Let
$g_b(X_{n/2})=\phi'(b)\in \Gamma'$, $b\in \mathbb{F}_2^{n/2}$. When
$g_b\in \Gamma'_0$ and $g_b=c\cdot X_{n/2}\neq g'$, we have
\begin{eqnarray}
W_{g_{b}}(\alpha) = \left\{ \begin{array}{ll}
0 & \textrm{if $\alpha\neq c$}\\
2^{n/2} & \textrm{if $\alpha= c$}.
\end{array} \right.
\end{eqnarray}
From (\ref{36}), if $\alpha \neq c'$ and $(\alpha_{i_{1}}, \cdots,
\alpha_{i_{m+1}})= (1 \cdots 1)$, then $\alpha\neq c$. Obviously,
$\Gamma'_0$ is a set of disjoint spectra functions. So we have
\begin{align}
\sum_{\phi(b)\in \Gamma'_0 \atop b\in
\mathbb{F}_2^{n/2}}(-1)^{\beta\cdot b}W_{g_b}(\alpha)\in \{0, \pm
2^{m+2}, \pm(2^{n/2}-2^{m+2}), \pm 2^{n/2}\}.
\end{align}
Let $A_k=\Gamma_k \cap \{\phi(b) \ | \  b\in \mathbb{F}_2^{n/2} \}.$
Similarly to the proof of Theorem 1, for any $(\beta,\alpha)\in
\mathbb{F}_2^{n}$, we have
$$|W_{f'}(\beta,\alpha)|\leq 2^{n/2}+\sum_{k=1}^s a_k\cdot 2^{n/2-k}$$
where
$$
a_k = \left\{ \begin{array}{ll}
0 & \textrm{if $A_k=\emptyset$}\\
1 & \textrm{if $A_k\neq \emptyset$}.
\end{array} \right.
$$
From (\ref{3}), Inequality (39) holds and $f'$ is obviously almost
optimal. For the existence of $g'$, $deg(f')=n-m-1$. $\Box$

{\emph{Remark 3:}}

1) Apparently, the idea above to obtain degree-optimized resilient
functions is firstly considered by Pasalic \cite{Pasalic}.

2)  A long list of input instances and the corresponding
cryptographic parameters can be found in Table 1 and Table 2. In
Table 2, the entries with ``*" represent the functions that can not
be degree optimized via Construction 2 on the premise of that
$N_{f'}=N_f$.

\begin{table*}[tbp]
\begin{center}
\caption{Existence of Almost Optimal $(n, m,n-m-1, N_{f'})$
functions ($1\leq m\leq 4$)}
\begin{tabular}{|c|c|c|c|c}\hline
%\backslashbox{}
 $m$ &  \multicolumn{2}{|c|}{$n$}              &~$N_{f'}  $

 \\\hline
  &&$12\leq n\leq 20$     &$2^{n-1}-2^{n/2-1}-2^{n/4+1}-4$
 \\&&$24\leq n\leq 112$     & $2^{n-1}-2^{n/2-1}-2^{n/4+2}-4$
 \\&&$116\leq n\leq 132$    & $2^{n-1}-2^{n/2-1}-2^{n/4+2}-2^{n/4+1}-4$
 \\&$n\equiv 0\atop{(mod 4)}$&$n=136$                & $2^{n-1}-2^{n/2-1}-2^{n/4+2}-2^{n/4+1}-2^{n/4}-4$
 \\&&$140\leq n\leq 492$     & $2^{n-1}-2^{n/2-1}-2^{n/4+3}-4$
 \\&&$496\leq n\leq 512$     & $2^{n-1}-2^{n/2-1}-2^{n/4+3}-2^{n/4+1}-4$
 %\\&&$n=516$                & $2^{n-1}-2^{n/2-1}-2^{n/4+3}-2^{n/4+1}-2^{n/4}-4$
% \\&&$520\leq n\leq 604$    & $2^{n-1}-2^{n/2-1}-2^{n/4+3}-2^{n/4+2}-4$
% \\1&&$n=608$               & $2^{n-1}-2^{n/2-1}-2^{n/4+3}-2^{n/4+2}-2^{n/4}-4$
% \\&&$612\leq n\leq 628$    & $2^{n-1}-2^{n/2-1}-2^{n/4+3}-2^{n/4+2}-2^{n/4+1}-4$
% \\&&$632\leq n\leq 2024$   & $2^{n-1}-2^{n/2-1}-2^{n/4+4}-4$
 \\\cline{2-4}

 & &$14\leq n\leq 50$    & $2^{n-1}-2^{n/2-1}-2^{{(n+6)}/4}-4$
 \\&&$54\leq n\leq 58$    & $2^{n-1}-2^{n/2-1}-2^{{(n+6)}/4}-2^{{(n+2)}/4}-4$
 \\&&$62\leq n\leq 238$   & $2^{n-1}-2^{n/2-1}-2^{{(n+10)}/4}-4$
 \\&$n\equiv 2\atop{(mod 4)}$
  &$242\leq n\leq 246$    & $2^{n-1}-2^{n/2-1}-2^{{(n+10)}/4}-2^{{(n+2)}/4}-4$
 \\&&$250\leq n\leq 290$   & $2^{n-1}-2^{n/2-1}-2^{{(n+10)}/4}-2^{{(n+6)}/4}-4$
 \\&&$294\leq n\leq 298$    & $2^{n-1}-2^{n/2-1}-2^{{(n+10)}/4}-2^{{(n+6)}/4}-2^{{(n+2)}/4}-4$
% \\&&$302\leq n\leq 1002$  & $2^{n-1}-2^{n/2-1}-2^{{(n+14)}/4}-4$
 \\\hline

  &&$n=16$                 &$2^{n-1}-2^{n/2-1}-2^{n/4+2}-8$
 \\&&$20\leq n\leq 40$        &$2^{n-1}-2^{n/2-1}-2^{n/4+3}-8$
 \\&&$n=44$                   &$2^{n-1}-2^{n/2-1}-2^{n/4+3}-2^{n/4+2}-8$
 \\&$n\equiv 0\atop{(mod 4)}$
  &$48\leq n\leq 84$     & $2^{n-1}-2^{n/2-1}-2^{n/4+4}-8$
 \\2&&$n=88$                  & $2^{n-1}-2^{n/2-1}-2^{n/4+4}-2^{n/4+2}-8$
 \\&&$92\leq n\leq 96$         & $2^{n-1}-2^{n/2-1}-2^{n/4+4}-2^{n/4+3}-8$
 \\&&$100\leq n\leq 176$      & $2^{n-1}-2^{n/2-1}-2^{n/4+5}-8$
%&\\&&$n=180$                         & $2^{n-1}-2^{n/2-1}-2^{n/4+5}-2^{n/4+3}$
%&\\&&$n=184$                         & $2^{n-1}-2^{n/2-1}-2^{n/4+5}-2^{n/4+3}-2^{n/4+2}-2^{n/4+1}$
%&\\&{$\displaystyle{m=2} \atop {n\equiv 0\ (mod 4)}$}&$188\leq n\leq 196$          &$3n/4-5$     & $2^{n-1}-2^{n/2-1}-2^{n/4+5}-2^{n/4+4}$
%&\\&&$n=200$                         & $2^{n-1}-2^{n/2-1}-2^{n/4+5}-2^{n/4+4}-2^{n/4+3}$
%&\\&&$204\leq n\leq 356$             & $2^{n-1}-2^{n/2-1}-2^{n/4+6}$
%&\\&&$n=360$                         & $2^{n-1}-2^{n/2-1}-2^{n/4+6}-2^{n/4+2}$
%&\\&&$364\leq n\leq 368$              & $2^{n-1}-2^{n/2-1}-2^{n/4+6}-2^{n/4+4}$
%&\\&&$n=372$                          & $2^{n-1}-2^{n/2-1}-2^{n/4+6}-2^{n/4+4}-2^{n/4+3}$
%&\\&&$376\leq n\leq 400$              & $2^{n-1}-2^{n/2-1}-2^{n/4+6}-2^{n/4+5}$
%&\\&&$n=404$                          & $2^{n-1}-2^{n/2-1}-2^{n/4+6}-2^{n/4+5}-2^{n/4+3}$
%&\\&&$n=408$                          & $2^{n-1}-2^{n/2-1}-2^{n/4+6}-2^{n/4+5}-2^{n/4+4}$
%&\\&&$n=412$                           & $2^{n-1}-2^{n/2-1}-2^{n/4+6}-2^{n/4+5}-2^{n/4+4}-2^{n/4+3}$
%&\\&&$416\leq n\leq 720$         &$2^{n-1}-2^{n/2-1}-2^{n/4+7}$ &
\\\cline{2-4}

  &&$18\leq n\leq 26$                  &$2^{n-1}-2^{n/2-1}-2^{(n+10)/4}-8$
 \\&$n\equiv 2\atop{(mod 4)} $&$30\leq n\leq 58$ &$2^{n-1}-2^{n/2-1}-2^{(n+14)/4}-8$
\\&&$62\leq n\leq 66$                 &$2^{n-1}-2^{n/2-1}-2^{(n+14)/4}-2^{(n+10)/4}-8$
\\&&$70\leq n\leq 122$                 &$2^{n-1}-2^{n/2-1}-2^{(n+18)/4}-8$

\\\hline

  &&$n=20$                          &$2^{n-1}-2^{n/2-1}-2^{n/4+3}-2^{n/4+2}-16$
 \\&&$24\leq n\leq 32$              &$2^{n-1}-2^{n/2-1}-2^{n/4+4}-16$
\\&&$n=36$                           &$2^{n-1}-2^{n/2-1}-2^{n/4+4}-2^{n/4+3}-16$
\\&&$40\leq n\leq 56$              &$2^{n-1}-2^{n/2-1}-2^{n/4+5}-16-16$
\\&$n\equiv 0\atop{(mod 4)}$
  &$n=60$                           &$2^{n-1}-2^{n/2-1}-2^{n/4+5}-2^{n/4+4}-16$
\\&&$64\leq n\leq88$                 &$2^{n-1}-2^{n/2-1}-2^{n/4+6}-16$
\\&&$n=92$                          &$2^{n-1}-2^{n/2-1}-2^{n/4+6}-2^{n/4+4}-16$
\\3&&$n=96$                           &$2^{n-1}-2^{n/2-1}-2^{n/4+6}-2^{n/4+5}-16$
\\&&$100\leq n\leq 144$             &$2^{n-1}-2^{n/2-1}-2^{n/4+7}-16$
\\\cline{2-4}

 & &$22\leq n\leq 26$            &$2^{n-1}-2^{n/2-1}-2^{(n+14)/4}-16$
\\&&$30\leq n\leq 42$             &$2^{n-1}-2^{n/2-1}-2^{{(n+18)}/4}-16$
\\&    &$n=46$                    &$2^{n-1}-2^{n/2-1}-2^{{(n+18)}/4}-2^{{(n+14)}/4}-16$
\\&$n\equiv 2\atop{(mod 4)}$
      &$52\leq n\leq 70$       &$2^{n-1}-2^{n/2-1}-2^{{(n+22)}/4}-16$
\\&    &$n=74$               &$2^{n-1}-2^{n/2-1}-2^{{(n+22)}/4}-2^{{(n+18)}/4}-16$
\\&    &$n=78$               &$2^{n-1}-2^{n/2-1}-2^{{(n+22)}/4}-2^{{(n+18)}/4}-2^{{(n+14)}/4}-16$
\\&&$82\leq n\leq 114$        &$2^{n-1}-2^{n/2-1}-2^{{(n+26)}/4}-16$
\\\hline

 &&$28\leq n\leq 32$        &$2^{n-1}-2^{n/2-1}-2^{n/4+5}-32$
\\&&$36\leq n\leq 48$     &$2^{n-1}-2^{n/2-1}-2^{n/4+6}-32$
\\&&$n=52$               &$2^{n-1}-2^{n/2-1}-2^{n/4+6}-2^{n/4+5}-32$
\\&$n\equiv 0\atop{(mod 4)}$&$56\leq n\leq 68$  & $2^{n-1}-2^{n/2-1}-2^{n/4+7}-32$
\\&&$n=72$              &$2^{n-1}-2^{n/2-1}-2^{n/4+7}-2^{n/4+5}-2^{n/4+4}-32$
\\&&$76\leq n\leq 100$  & $2^{n-1}-2^{n/2-1}-2^{n/4+8}-32$
\\\cline{2-4}

     & &$n=26$               &$2^{n-1}-2^{n/2-1}-2^{{(n+18)}/4}-32$
 \\4 &&$30\leq n\leq 38$     &$2^{n-1}-2^{n/2-1}-2^{{(n+22)}/4}-32$
 \\&    &$n=42$              &$2^{n-1}-2^{n/2-1}-2^{{(n+22)}/4}-2^{{(n+18)}/4}-32$
\\&$n\equiv 2\atop{(mod 4)}$ &$46\leq n\leq 58$     &$2^{n-1}-2^{n/2-1}-2^{{(n+26)}/4}-32$
\\&&$n=62$                   &$2^{n-1}-2^{n/2-1}-2^{{(n+26)}/4}-2^{{(n+22)}/4}-32$
\\&&$66\leq n\leq 82$        &$2^{n-1}-2^{n/2-1}-2^{{(n+30)}/4}-32$
\\&&$n=86$                   &${ 2^{n-1}-2^{n/2-1}-2^{{(n+30)}/4}-2^{{(n+22)}/4}-2^{{(n+18)}/4}-2^{(n+14)/4}-32}$
\\&&$n=90$                   &$2^{n-1}-2^{n/2-1}-2^{{(n+30)}/4}-2^{{(n+26)}/4}-2^{{(n+22)}/4}-32$
\\&&$94\leq n\leq 118$       &$2^{n-1}-2^{n/2-1}-2^{{(n+34)}/4}-32$
\\\hline

\end{tabular}
\end{center}
\end{table*}

\begin{table*}[tbp]
\begin{center}
 \caption{$(n, m, n-m-1, N_{f'})$ functions ($m\geq 5$)which were not known earlier}
\begin{tabular}{|c|c|c}\hline
%\backslashbox{}
 $(30, 5,24,2^{29}-2^{14}-2^{13})$& $(36,5, 30,2^{35}-2^{17}-2^{15}-2^{6})^*$
 \\$(38, 5,32,2^{37}-2^{18}-2^{16})$ &$(42,5,36,2^{41}-2^{20}-2^{17}-2^{14}-2^{6})^*$
 \\$(44, 5, 38, 2^{43}-2^{21}-2^{18}-2^{6})^*$ &$(48,5,42, 2^{47}-2^{23}-2^{19}-2^{6})^*$
 \\$(54,5,48, 2^{53}-2^{26}-2^{21}-2^{6})^*$ & $(58,5,52, 2^{57}-2^{28}-2^{22}-2^{21}-2^{6})^*$
 \\ $(60,5,54, 2^{59}-2^{29}-2^{23}-2^{6})^*$ &$(64,5,48, 2^{63}-2^{31}-2^{24}-2^{6})^*$
 \\ $(70,5,64, 2^{69}-2^{34}-2^{26}-2^{6})^*$      & $(74,5,68, 2^{73}-2^{36}-2^{27}-2^{24}-2^{6})^*$
 \\$(76,5,70, 2^{75}-2^{37}-2^{28}-2^{6})^*$&$(80,5,74, 2^{79}-2^{39}-2^{29}-2^{6})^*$
 \\$(84,5,78, 2^{83}-2^{41}-2^{30}-2^{6})^*$ &$(88,5,82, 2^{87}-2^{43}-2^{31}-2^{30}-2^{6})^*$
 \\$(90,5,84, 2^{89}-2^{44}-2^{32}-2^{6})^*$      & $(94,5,88, 2^{93}-2^{46}-2^{33}-2^{6})^*$
 \\$(98,5,92, 2^{97}-2^{48}-2^{34}-2^{32}-2^{6})^*$& $(100,5,94, 2^{99}-2^{49}-2^{35}-2^{6})^*$
% \\$(200,5,194, 2^{199}-2^{99}-2^{62}-2^{61}-2^{6})^*$&

%$(1000,5,994, 2^{999}-2^{499}-2^{268}-2^{6})^*$
%$(40, 5, 34,2^{39}-2^{19}-2^{17})$
% $(32,5,26,2^{31}-2^{15}-2^{14})$
% $(46, 5, 40,2^{45}-2^{22}-2^{19}-2^{6})^*$
%$(50,5,44, 2^{49}-2^{24}-2^{20}-2^{6})^*$&
% $(52,5,46,2^{51}-2^{25}-2^{21}-2^{6})^*$
% $(86,5,80,2^{85}-2^{42}-2^{31}-2^{6})^*$
% $(62,5,56,2^{61}-2^{30}-2^{24}-2^{6})^*$
% $(66,5,60,2^{65}-2^{32}-2^{25}-2^{6})^*$
% $(68,5,62,2^{67}-2^{33}-2^{26}-2^{6})^*$
% $(72,5,66,2^{71}-2^{35}-2^{27}-2^{6})^*$
% $(78,5,72,2^{77}-2^{38}-2^{29}-2^{6})^*$
% $(82,5,76,2^{81}-2^{40}-2^{30}-2^{6})^*$
% $(56,5,50,2^{55}-2^{27}-2^{22}-2^{6})^*$

 \\\hline
 $(34,6,27,2^{23}-2^{16}-2^{15})$             &$(40, 6,33,2^{39}-2^{19}-2^{17}-2^{16}-2^{7})^*$
 \\$(42, 6, 35,2^{41}-2^{20}-2^{18})$   & $(48,6,41, 2^{47}-2^{23}-2^{20}-2^{7})^*$
 \\$(52,6,45, 2^{51}-2^{25}-2^{22})$ &$(54,6,47, 2^{53}-2^{26}-2^{22}-2^{7})^*$
 \\$(60,6,53, 2^{59}-2^{29}-2^{24}-2^{7})^*$ &$(64,6,47, 2^{63}-2^{31}-2^{25}-2^{24}-2^{7})^*$
 \\$(66,6,59, 2^{65}-2^{32}-2^{26}-2^{7})^*$ & $(70,6,63, 2^{69}-2^{34}-2^{27}-2^{7})^*$
 \\$(76,6,69,2^{75}-2^{37}-2^{29}-2^{7})^*$ &$(80,6,73, 2^{79}-2^{39}-2^{30}-2^{29}-2^{7})^*$
 \\ $(82,6,75,2^{81}-2^{40}-2^{31}-2^{7})^*$  &$(86,6,79,  2^{85}-2^{42}-2^{32}-2^{7})^*$
 \\$(90,6,83, 2^{89}-2^{44}-2^{33}-2^{32}-2^{7})^*$ & $(92,6,85,2^{91}-2^{45}-2^{34}-2^{7})^*$
 \\$(96,6,89,  2^{95}-2^{47}-2^{35}-2^{7})^*$ &$(100,6,93, 2^{99}-2^{49}-2^{36}-2^{35}-2^{7})^*$

%\\& $(200,6,193, 2^{199}-2^{199}-2^{164}-2^{163}-2^{7})^*$
% $(36,6,29,2^{35}-2^{17}-2^{16})$&
% $(38,6,31, 2^{37}-2^{18}-2^{17})$
% $(50,6,43, 2^{49}-2^{24}-2^{21}-2^{7})^*$ &
% $(44, 6, 37,2^{43}-2^{21}-2^{19})$
% $(56,6,49, 2^{55}-2^{27}-2^{23}-2^{7})^*$&
% $(62,6,55, 2^{61}-2^{30}-2^{25}-2^{7})^*$&
% $(78,6,71, 2^{77}-2^{38}-2^{30}-2^{7})^*$
% $(58,6,51,2^{57}-2^{28}-2^{24}-2^{7})^*$
% $(72,6,65, 2^{71}-2^{35}-2^{28}-2^{7})^*$&
% $(74,6,67, 2^{73}-2^{36}-2^{29}-2^{7})^*$&
% $(84,6,77, 2^{83}-2^{41}-2^{32}-2^{7})^*$
% $(68,6,61, 2^{67}-2^{33}-2^{27}-2^{7})^*$&

 \\\hline
  $(38,7,30,2^{37}-2^{18}-2^{17}-2^{16})$& $(40, 7, 32,2^{39}-2^{19}-2^{18})$
  \\$(46, 7, 38, 2^{45}-2^{22}-2^{20})$ &$(48,7,40, 2^{47}-2^{23}-2^{21})$
  \\$(52,7,44, 2^{51}-2^{25}-2^{22}-2^{21}-2^{8})^*$ &$(54,7,46, 2^{53}-2^{26}-2^{23})$
 \\ $(58,7,50, 2^{57}-2^{28}-2^{24}-2^{23}-2^{8})^*$&$(60,7,52, 2^{59}-2^{29}-2^{25}-2^{8})^*$
 \\$(64,7,46, 2^{63}-2^{31}-2^{26}-2^{25}-2^{8})^*$ &$(66,7,58, 2^{65}-2^{32}-2^{27}-2^{8})^*$
 \\$(70,7,62, 2^{69}-2^{34}-2^{28}-2^{27}-2^{8})^*$&$(72,7,64, 2^{71}-2^{35}-2^{29}-2^{8})^*$
 \\ $(76,7,68, 2^{73}-2^{37}-2^{30}-2^{8})^*$& $(78,7,70, 2^{77}-2^{38}-2^{31}-2^{8})^*$
 \\$(82,7,74, 2^{81}-2^{40}-2^{32}-2^{8})^*$ &$(86,7,78, 2^{85}-2^{42}-2^{33}-2^{32}-2^{8})^*$
 \\$(88,7,80, 2^{87}-2^{43}-2^{34}-2^{8})^*$& $(92,7,84,2^{91}-2^{45}-2^{35}-2^{8})^*$
 \\$(98,7,90, 2^{97}-2^{48}-2^{37}-2^{8})^*$&$(100,7,92, 2^{99}-2^{49}-2^{38}-2^{8})^*$
 %\\$(200,7,182,2^{199}-2^{99}-2^{66}-2^{63}-2^{8})^*$

%
% $(84,7,76,2^{83}-2^{41}-2^{33}-2^{8})^*$&
% $(42, 7, 34,2^{41}-2^{20}-2^{19})$
% $(50,7,42, 2^{49}-2^{24}-2^{22})$
% $(56,7,48,2^{55}-2^{27}-2^{24})$
% $(62,7,54, 2^{61}-2^{30}-2^{26}-2^{8})^*$
% $(80,7,72, 2^{79}-2^{39}-2^{32}-2^{8})^*$
% $(74,7,66,(2^{73}-2^{36}-2^{30}-2^{8})^*$
% $(68,7,60,2^{67}-2^{33}-2^{28}-2^{8})^*$

  \\\hline
  $(42, 8, 33, 2^{41}-2^{20}-2^{19}-2^{18})$  &$(44, 8, 35, 2^{43}-2^{21}-2^{20})$
  \\$(50,8,41, 2^{49}-2^{24}-2^{22}-2^{21})$  & $(52,8,43, 2^{51}-2^{25}-2^{23})$
  \\$(58,8,49, 2^{57}-2^{28}-2^{25}-2^{9})^*$  & $(64,8,45, 2^{63}-2^{31}-2^{27}-2^{9})^*$
 \\ $(68,8,59, 2^{67}-2^{33}-2^{25}-2^{29})$  &$(70,8,61, 2^{69}-2^{34}-2^{29}-2^{27}-2^{9})^*$
 \\$(72,8,63, 2^{71}-2^{35}-2^{30}-2^{9})^*$ &$(76,8,67,2^{75}-2^{37}-2^{31}-2^{28}-2^{9})^*$
 \\$(78,8,69, 2^{77}-2^{38}-2^{32}-2^{9})^*$          &$(82,8,73,2^{81}-2^{40}-2^{33}-2^{9})^*$
 \\ $(88,8,79,2^{87}-2^{43}-2^{35}-2^{9})^*$          &$(92,8,83, 2^{91}-2^{45}-2^{36}-2^{35}-2^{9})^*$
 \\$(94,8,85,2^{93}-2^{46}-2^{37}-2^{9})^*$   & $(98,8,89,2^{97}-2^{48}-2^{38}-2^{36}-2^{9})^*$
 \\$(100,8,91, 2^{99}-2^{49}-2^{39}-2^{9})^*$ &$(200,8,191, 2^{199}-2^{99}-2^{68}-2^{9})^*$

% $(54,8,45, 2^{53}-2^{26}-2^{24})$
% $(62,8,53, 2^{61}-2^{30}-2^{27})$
% $(74,8,65, 2^{73}-2^{36}-2^{31})$
% $(80,8,71, 2^{79}-2^{39}-2^{33})$
% $(46, 8, 37, 2^{45}-2^{22}-2^{21})$
% $(84,8,75,2^{83}-2^{41}-2^{34})$
% $(66,8,57, 2^{65}-2^{32}-2^{28})$
% $(60,8,51,2^{59}-2^{29}-2^{26})$
% \\$(86,8,77, 2^{85}-2^{42}-2^{35})$& &
 \\\hline
 $(46,9,36,{2^{45}-2^{22}-2^{21}-2^{20}-2^{19}-2^{10}})^*$&$(48,9,38, 2^{47}-2^{23}-2^{22})$
 \\$(54,9,44, 2^{53}-2^{26}-2^{24}-2^{23}-2^{22})$  &$(56,9,46, 2^{55}-2^{27}-2^{25})$
 \\$(62,9,52, 2^{61}-2^{30}-2^{27}-2^{26})$  &$(64,9,44, 2^{63}-2^{31}-2^{28})$
 \\$(68,9,58,{ 2^{67}-2^{33}-2^{29}-2^{28}-2^{27}-2^{10}})^*$ & $(70,9,60, 2^{69}-2^{34}-2^{30}-2^{10})^*$
 \\$(74,9,64, 2^{73}-2^{36}-2^{32}$  &$(76,9,66, 2^{75}-2^{37}-2^{32}-2^{10})^*$
 \\ $(80,9,70, 2^{79}-2^{39}-2^{34}-2^{10})^*$  &$(82,9,72, 2^{81}-2^{40}-2^{34}-2^{10})^*$
 \\$(88,9,78, 2^{87}-2^{43}-2^{36}-2^{10})^*$      & $(94,9,84, 2^{93}-2^{46}-2^{38}-2^{10})^*$
 \\$(98,9,88, 2^{97}-2^{48}-2^{39}-2^{38}-2^{10})^*$  &$(100,9,90, 2^{99}-2^{49}-2^{40}-2^{10})^*$
% \\$(200,9,190, 2^{199}-2^{99}-2^{70}-2^{10})^*$& $(300,9,290, {\scriptstyle2^{299}-2^{159}-2^{107}-2^{106}-2^{104}-2^{10}})^*$

% $(58,9,48, 2^{57}-2^{28}-2^{26})$
% $(84,9,74, 2^{83}-2^{41}-2^{35})$
% $(72,9,62, 2^{71}-2^{35}-2^{31})$
% $(78,9,68,2^{77}-2^{38}-2^{33})$
% $(86,9,76,2^{85}-2^{42}-2^{36})$
% $(66,9,56, 2^{65}-2^{32}-2^{29})$

 \\\hline
   $(52,10,41, 2^{51}-2^{25}-2^{24})$&$(60,10,49, 2^{59}-2^{29}-2^{27})$
   \\${\scriptstyle(66,10,55,2^{65}-2^{32}-2^{29}-2^{28}-2^{27}-2^{26}-2^{25}-2^{11})}^*$  &$(68,10,57, 2^{67}-2^{33}-2^{30})$
   \\$(74,10,63,{2^{73}-2^{36}-2^{32}-2^{30}-2^{11}})^*$         &$(76,10,65, 2^{75}-2^{37}-2^{33})$
 \\$(80,10,69,{2^{79}-2^{39}-2^{34}-2^{33}-2^{11}})^*$& $(82,10,71, 2^{81}-2^{40}-2^{35}-2^{11})^*$
 \\$(84,10,73, 2^{83}-2^{41}-2^{36}-2^{11})^*$  &$(86,10,75,{ 2^{85}-2^{42}-2^{36}-2^{35}-2^{34}-2^{11}})^*$
 \\$(88,10,77, 2^{87}-2^{43}-2^{37}-2^{11})^*$ &$(92,10,81,{2^{91}-2^{45}-2^{38}-2^{37}-2^{36}-2^{35}-2^{11}})^*$
 \\$(94,10,83, 2^{93}-2^{46}-2^{39}-2^{11})^*$& $(98,10,87, {2^{97}-2^{48}-2^{40}-2^{39}-2^{38}-2^{11}})^*$
 \\$(100,10,89, 2^{99}-2^{49}-2^{41}-2^{11})^*$& $(500,10,489, 2^{499}-2^{249}-2^{153}-2^{11})^*$

 \\\hline  $(100,21,78, 2^{99}-2^{49}-2^{48})$& $(200,45,154, 2^{199}-2^{99}-2^{98})$
 \\ $(184,38,145, 2^{183}-2^{91}-2^{89}-2^{87}-2^{86})$ & $(516,116,399, 2^{515}-2^{255}-2^{253})$
 \\ $(832,200,631, 2^{831}-2^{415}-2^{414}-2^{413})$& ${(10000,2475,7524, 2^{9999}-2^{4999}-2^{4998}-2^{4997}-2^{4996})}$
 \\\hline
\end{tabular}
\end{center}
\end{table*}

\section{Improved Version of the Main Construction}

Both constant functions and balanced Boolean functions are regarded
as $0$-resilient functions. The Boolean functions that are neither
balanced nor correlation-immune are regarded as $(-1)$-resilient
functions (e.g. bent functions).

{\emph{Lemma 4:}} With the same notation as in the Definition 4, if
$h_c$ is a $v$-resilient function, then $g_c$ is a
$(wt(c)+v)$-resilient function.

{\emph{Proof:}} Let $\alpha\in \mathbb{F}_2^p$ and $l=c\cdot X'_t$.
It is not difficult to deduce that $
W_{g_c}(\alpha)=W_{l}(\alpha_{i_1}, \cdots, \alpha_{i_t})\cdot
W_{h_c}(\alpha_{i_{t+1}}, \cdots, \alpha_{i_p}).$ When
$wt(\alpha_{i_1}, \cdots, \alpha_{i_t})< wt(c)$,
$W_{l}(\alpha_{i_1}, \cdots, \alpha_{i_t})=0$. From Lemma 1, for
$h_c$ is a $v$-resilient function,  we have
\begin{eqnarray*}
~~W_{h_c}(\alpha_{i_{t+1}}, \cdots, \alpha_{i_p})=0, ~~\textrm{for
$wt(\alpha_{i_{t+1}}, \cdots, \alpha_{i_p})\leq v$}.
\end{eqnarray*}
Obviously, $W_{g_c}(\alpha)=0$ when $wt(\alpha)\leq wt(c)+v$. From
Lemma 1, $g_c$ is a $(wt(c)+v)$-resilient function. $\Box$

{\emph{Construction 3:}} Let $n\geq 12$ be an  even number, $m$ be a
positive number, $e_k$ be a nonnegative number with $0\leq e_k\leq
k+1$, and $a_k\in \mathbb{F}_2$ ($k=1, \cdots, \lfloor n/4 \rfloor$)
such that
\begin{align}
\sum_{i=m+1}^{n/2} {{n/2}\choose i} + \sum_{k=1}^{\lfloor
n/4\rfloor}\left(a_k \cdot \sum_{j=m-e_k+1}^{n/2-2k}
{{n/2-2k}\choose j}\right)\geq 2^{n/2}.
\end{align}
 Let $X_{n/2}=(x_1,\cdots, x_{n/2})\in
\mathbb{F}_2^{n/2}$, $X'_{t}=(x_1,\cdots, x_{t})\in
\mathbb{F}_2^{t}$ and $X''_{2k}=(x_{t+1},\cdots, x_{n/2})\in
\mathbb{F}_2^{2k}$, where $t+2k=n/2$. Let
\begin{eqnarray}
\Omega_0=\{c \cdot X_{n/2}\ | \ c\in \mathbb{F}_2^{n/2}, \ wt(c)>m
\}.
\end{eqnarray}  For
$1\leq k\leq \lfloor n/4\rfloor$ and $0 \leq e_{k}\leq m+1$, let
$R_k$ be a nonempty set of nonlinear $(2k, e_k-1, -, N_{h_k})$
functions with high nonlinearity and
\begin{align}
\Omega_k=\{c\cdot X'_{t}\oplus h_c(X''_{2k})~|~c\in
\mathbb{F}_2^t,wt(c)> m-e_k\}
\end{align}
where $h_c\in R_k$. Set
\begin{eqnarray}
\Omega=\bigcup_{k=0}^{\lfloor n/4\rfloor}\Omega_k.
\end{eqnarray}

Denote by $\varphi$ any injective mapping from $\mathbb{F}_2^{n/2}$
to $\Omega$  such that there exists an $(n/2, m, n/2-m-1, N_{g_b})$
function $g_b \in \Omega$ with $\varphi^{-1}(g_b)\neq \emptyset$ .
We construct the function $f\in \mathcal {B}_n$ as follows:
\begin{eqnarray}
f(Y_{n/2},X_{n/2})= \bigoplus_{b\in
\mathbb{F}_2^{n/2}}Y_{n/2}^b\cdot \varphi(b)
\end{eqnarray}

{ \emph{Theorem 3:}} If $f\in \mathcal {B}_n$ is proposed by
Construction 3, then $f$ is an almost optimal $(n,m,n-m-1,N_f)$
function with
\begin{align}\label{49}
N_f\geq 2^{n-1}-2^{n/2-1}-\sum_{k=1}^{\lfloor n/4\rfloor}(a_k\cdot
2^{n/2-2k} \cdot (2^{2k-1}-N_{h_k})).
\end{align}

{ \emph{Proof:}} For any $(\beta,\alpha)\in \mathbb{F}_2^{n/2}\times
\mathbb{F}_2^{n/2}$ we have
\begin{eqnarray}
 W_f(\beta,\alpha)
&=& \sum_{k=0}^s \sum_{\phi(b)\in \Gamma_k \atop b\in
\mathbb{F}_2^{n/2}}(-1)^{\beta\cdot b}W_{g_b}(\alpha)
\end{eqnarray}
Let
\begin{eqnarray}
A_k=\Omega_k \cap \{\phi(b) \ | \  b\in \mathbb{F}_2^{n/2} \}.
\end{eqnarray}
Note that each $\Omega_k$ ($k=0, 1, \cdots, \lfloor n/4\rfloor$) is
a set of disjoint spectra functions. Similarly to the proof of
Theorem 1, we obtain
\begin{eqnarray}
|W_f(\beta,\alpha)|\leq 2^{n/2}+\sum_{k=1}^s a_k\cdot
2^{n/2-2k}\cdot (2^{2k}-2N_{h_k})
\end{eqnarray}
where
$$
a_k = \left\{ \begin{array}{ll}
0 & \textrm{if $A_k=\emptyset$}\\
1 & \textrm{if $A_k\neq \emptyset$}.
\end{array} \right.
$$
From (\ref{3}), Inequality (\ref{49}) holds.

From Lemma 4, all the functions in $\Omega$ are $m$-resilient
functions. Due to Lemma 2, $f$ is an $m$-resilient function. For the
existence of a degree-optimized function $g_b \in \Omega$ with
$\varphi^{-1}(g_b)\neq \emptyset$, we have $deg(f)=n-m-1$. $\Box$

Fixing $n$ and $m$, we can also obtain lots of degree-optimized
resilient functions whose nonlinearity are better than that of
functions constructed by Construction 1. See the following example.

{\emph{Example 2:}} It is possible to construct a $(28, 1, 26,
2^{27}-2^{13}-2^{8}-2^{6})$ function. Let
$$
\Omega_0 = \{c\cdot X'_{t}~ |~c\in \mathbb{F}_2^{14},~wt(c)\geq 2 \}
$$
and
$$
\Omega_5=\{c\cdot X'_{4}\oplus h_c(X''_{10})~ |~c\in
\mathbb{F}_2^{4},~h_c\in R_5\}
$$
where $R_5$ is a nonempty set of $(10, 1, 8, 492)$ functions
\cite{Kavut}.  Note that $|\Omega_0|=16369$, $|\Omega_5| = 16$. For
$16369+16>2^{14}$, it is possible to select $2^{14}$ many
$14$-variable $1$-resilient functions from $\Omega_0\cup \Omega_5$.
We  concatenate these functions and obtain a $(28, 1, 26,
2^{27}-2^{13}-2^{8}-2^{6})$ function.

 Similarly,  one can obtain the following resilient functions:  $(36, 3, 32, 2^{35}-2^{17}-2^{13})$, $(42, 5, 36,
 2^{41}-2^{20}-2^{17}-2^{14})$,  $(66, 10, 55, 2^{65}-2^{32}-2^{29}-2^{28}-2^{27}-2^{26}-2^{25})$,
 $(86, 4, 81, 2^{85}-2^{42}-2^{29}-2^{27}-2^{26}-2^{18}-2^{14}-2^{13}-2^{5})$,
 etc.

\section{Conclusion and an Open Problem}

In this paper, we  described a technique for constructing resilient
functions with good nonlinearity on large even number variables. As
a consequence, we obtained general constructions of functions which
were not known earlier.

 Sarkar and Maitra \cite{Sarkar-c} have
shown that the nonlinearity of any $(n, m, n-m-1, N_f)$ function
($m\leq n-2$) is divisible by $2^{m+2}$. And they have deduced the
following  result: If $n$ is even, and $m \leq n/2-2$, then $N_f\leq
2^{n-1}-2^{n/2-1}-2^{m+1}$. But we  suppose that this upper bound
could be improved. So we propose an open problem as follows:

Does there exist $n$-variable ($n$ even), $m$-resilient ($m\geq 0$)
functions with nonlinearity
 $> 2^{n-1}-2^{n/2-1}-2^{\lfloor n/4 \rfloor+m-1}$? If there
does, how to construct these functions?

{\emph{Conjecture:}} Let $n\geq 12$ be even and $m\leq n/2-2$. For
any $(n, m, - , N_f)$ function, the following inequality always
holds:
\begin{eqnarray}
N_f\leq 2^{n-1}-2^{n/2-1}-2^{\lfloor n/4 \rfloor+m-1}.
\end{eqnarray}


\begin{thebibliography}{11}

\bibitem{Camion} P. Camion, C. Carlet, P. Charpin, and N. Sendrier, ``On
correlation-immune functions," in Advances in Cryptology - CRYPTO'91
(Lecture Notes in Computer Sceince). Berlin, Germany:
Springer-Verlag, 1992, vol. 547, pp. 86-100.


\bibitem{Carlet} C. Carlet,  ``A larger class of cryptographic Boolean functions via a study of
the Maiorana-Mcfarland constructions," in Advances in Cryptology -
CRYPTO 2002 (Lecture Notes in Computer Sceince), Berlin, Germany:
Springer-Verlag, 2002, vol. 2442, pp. 549-564.


\bibitem{Chee} S. Chee, S. Lee, D. Lee, and S. H. Sung, ``On the correlation immune
functions and their nonlinearity," in Advances in Cryptology -
Asiacrypt'96 (Lecture Notes in Computer Sceince). Berlin, Germany:
Springer-Verlag, 1997, vol. 1163, pp. 232-243.

\bibitem{Clark} J. Clark, J. Jacob, S. Stepney, S. Maitra, and W. Millan, ``Evolving
Boolean functions satisfying multiple criteria," in Progress in
INDOCRYPT 2002 (Lecture Notes in Computer Science). Berlin, Germany:
Springer-Verlag, 2002, vol. 2551, pp. 246-259.

\bibitem{Dillon} J. F. Dillon, Elementary Hadamard difference set,
Ph.D. Thesis, University of Maryland, 1974.

\bibitem{Dobbertin} H. Dobbertin, ``Construction of bent functions and balanced Boolean
functions with high nonlinearity," in Workshop on Fast Software
Encryption (FES 1994) (Lecture Notes in Computer Science). Berlin,
Germany: Springer-Verlag, 1995, vol. 1008, pp. 61-74.




\bibitem{Fedorova} M. Fedorova and Y. V. Tarannikov, ``On the constructing of highly
nonlinear resilient Boolean functions by means of special matrices,"
in Progress in Cryptology - INDOCRYPT 2001 (Lecture Notes in
Computer Science). Berlin, Germany: Springer-Verlag, 2001, vol.
2247, pp. 254-266.




\bibitem{Kavut} S. Kavut, S. Maitra, and M. D. Y\"{u}cel, ``Search
for Boolean functions with excellent profiles in the rotation
symmetric class," IEEE Transations on Information Theory, vol. 53,
no. 5, pp. 1743-1751, 2007.

\bibitem{Khoo} K. Khoo, G. Gong, and H.-K. Lee, ``The rainbow attack
on stream ciphers based on Maiorana-McFarland functions," in Appiled
Cryptography and Network Security - ACNS 2006 (Lecture Notes in
Computer Sceince), Berlin, Germany: Springer-Verlag, 2006, vol.
3989, pp. 194-209.

\bibitem{MacWilliams} F. J. MacWilliams and N. J. A. Sloane, The Theory
of Error-Correcting Codes, Amsterdam, The Netherlands:
North-Holland, 1977.

\bibitem{Maitra-DAM} S. Maitra and E. Pasalic, ``A Maiorana-McFarland type
construction for resilient Boolean functions on variables ($n$ even)
with nonlinearity $>2^{n-1}-2^{n/2}+2^{n/2-2}$," Discrete Applied
Mathematics, vol. 154, pp. 357-369, 2006.

\bibitem{Maitra-IEEE} S. Maitra and E. Pasalic, ``Further constructions of resilient
Boolean functions with very high nonlinearity," IEEE Transations on
Information Theory, vol. 52, no. 5, pp. 2269-2270, 2006.

\bibitem{Merer} W. Meier and O. Staffelbach, ``Nonlinearity criteria
for cryptographic functions," in Advances in Cryptology -
EUROCRYPT'89 (Lecture Notes in Computer Sceince), Berlin, Germany:
Springer-Verlag, 1990, vol. 434, pp. 549-562.


\bibitem{Millan-i} W. Millan, A. Clark, and E. Dawson, ``An effective genetic algorithm
for finding highly nonlinear Boolean functions," in Proceedings of
the First International Conference on Information and Communication
Security (Lecture Notes in Computer Science). Berlin, Germany:
Springer-Verlag, 1997, vol. 1334, pp. 149-158.


\bibitem{Millan-a} W. Millan, A. Clark, and E. Dawson, ``Boolean function design using
hill climbing methods," in Proceedings of the 4th Australasian
Conference on Information Security and Privacy (Lecture Notes in
Computer Science). Berlin, Germany: Springer-Verlag, 1999, vol.
1587, pp. 1-11.

\bibitem{Millan-e} W. Millan, A. Clark, and E. Dawson. ``Heuristic design of
cryptographically strong balanced Boolean functions," in Advances in
Cryptology - EUROCRYPT'98 (Lecture Notes in Computer Science).
Berlin, Germany: Springer-Verlag, 1998, vol. 1403, pp. 489-499.

\bibitem{Pasalic} E. Pasalic, ``Maiorana-McFarland class: degree
optimization and algebraic properties," IEEE Transactions on
Information Theory, vol. 52, no.10, pp. 4581-4594, 2006.

\bibitem{Pasalic-wcc} E. Pasalic, S. Maitra, T. Johansson, and P. Sarkar, ``New
constructions of resilient and correlation immune Boolean functions
achieving upper bounds on nonlinearity," in Workshop on Coding and
Cryptography - WCC 2001, Paris, France, Jan. 8-12, 2001. Published
in Electronic Notes in Discrete Mathematics. Amsterdam, The
Netherlands: Elsevier Science, 2001, vol. 6, pp. 158-167.

\bibitem{Rothaus} O. S. Rothaus, On 'bent' functions, Journal of Combinatorial Theory,
Ser. A, vol. 20, pp. 300-305, 1976.

\bibitem{Saber} Z. Saber, M. F. Uddin, and A. Youssef,  ``On the existence of $(9, 3, 5,
240)$ resilient functions," IEEE Transations on Information Theory,
vol. 48, no. 7, pp. 1825-1834, 2002.

\bibitem{Sarkar-e} P. Sarkar and S. Maitra, ``Construction of nonlinear Boolean
functions with important cryptographic properties," in Advances in
Cryptology - EUROCRYPT 2000 (Lecture Notes in Computer Sceince),
Berlin, Germany: Springer-Verlag, 2000, vol. 1807, pp. 485-506.

\bibitem{Sarkar-ieee} P. Sarkar and S. Maitra, ``Efficient
implementation of cryptographically useful large Boolean functions,
IEEE Transactions on Computers, vol. 52, no. 4, pp. 410-417, 2003.



\bibitem{Sarkar-c} P. Sarkar and S. Maitra, ``Nonlinearity bounds and constructions of
resilient Boolean functions," in Advances in Cryptology - CRYPTO
2000 (Lecture Notes in Computer Sceince), Berlin, Germany:
Springer-Verlag, 2000, vol. 1880, pp. 515-532.


\bibitem{Sarkar} P. Sarkar and S. Maitra, ``Construction of nonlinear resilient
Boolean functions using small affine functions," IEEE Transactions
on Information Theory, vol. 50, no. 9, pp. 2185-2193, 2004.

\bibitem{Seberry} J. Seberry, X.-M. Zhang, and Y. Zheng, ``Nonlinearity
and propagation characteristics of balanced Boolean functions,"
Information and Computation, vol. 119, pp. 1-13, 1995.


\bibitem{Seberry-c} J. Seberry, X.-M. Zhang, and Y. Zheng, ``Nonlinearly balanced Boolean functions
and their propagation characteristics," in Advances in Cryptology -
CRYPTO'93 (Lecture Notes in Computer Sceince), Berlin, Germany:
Springer-Verlag, 1994, vol. 773, pp. 49-60.

\bibitem{Seberry-e} J. Seberry, X.-M. Zhang, and Y. Zheng, ``On constructions and
nonlinearity of correlation immune Boolean functions," in Advances
in Cryptology - EUROCRYPT'93 (Lecture Notes in Computer Sceince),
Berlin, Germany: Springer-Verlag, 1984, vol. 765, pp. 181-199.

\bibitem{Siegenthaler} T. Siegenthaler, ``Correlation-immunity of nonlinear combining
functions for cryptographic applications," IEEE Transactions on
Information Theory, vol. 30, no.5, pp. 776-780, 1984.

\bibitem{Tarannikov} Y. V. Tarannikov, ``On resilient Boolean functions with maximum
possible nonlinearity," in Progress in Cryptology - INDOCRYPT 2000
(Lecture Notes in Computer Science). Berlin, Germany: Springer
Verlag, 2000, vol. 1977, pp. 19-30.

\bibitem{Tarannikov-FSE}Y. V. Tarannikov, ``New constructions of resilient Boolean
functions with maximal nonlinearity," in Workshop on Fast Software
Encryption (FSE 2001) (Lecture Notes in Computer Science). Berlin,
Germany: Springer-Verlag, 2001, vol. 2355, pp. 66-77.


\bibitem{Xiao}GZ. Xiao and J. L. Massey, ``A spectral characterization of
correlation-immune combining functions," IEEE Transactions on
Information Theory, vol. 34, no. 3, pp. 569-571, 1988.




\end{thebibliography}
\end{document}